\documentstyle[12pt,pictex]{article}
\newcommand{\rf}[1]{(\ref{#1})}
\newcommand{\beq}{\begin{equation}}
\newcommand{\eeq}{\end{equation}}
\newcommand{\bea}{\begin{eqnarray}}
\newcommand{\eea}{\end{eqnarray}}

\renewcommand{\a}{\alpha}

\newcommand{\ep}{\varepsilon}

\newcommand{\oh}{\frac{1}{2}}

\newcommand{\cF}{{\cal F}}

\newcommand{\e}{\ep}
\newcommand{\F}{{\cal F}}
\newcommand{\T}{{\cal T}}

\newcommand{\ra}{\rangle}
\newcommand{\la}{\langle}

\begin{document}
\topmargin 0pt
\oddsidemargin 5mm
\headheight 0pt
\headsep 0pt
\topskip 9mm

\hfill    NBI-HE-96-9

\hfill    ITFA-96-3

\hfill LPTHE Orsay 96-15

\hfill January 1996

\begin{center}
\vspace{24pt}
{\large \bf RG flow in an exactly solvable model\\
 with fluctuating geometry}

\vspace{24pt}

{\sl J. Ambj\o rn }

\vspace{6pt}

The Niels Bohr Institute\\
Blegdamsvej 17, DK-2100 Copenhagen \O , Denmark\\         

\vspace{24pt}

{\sl P. Bialas}\footnote{Permanent address: Institute of Comp. Science,
Jagellonian University, ul. Nawojki 11, 30-072, Krak\'{o}w, Poland.}

\vspace{6pt}            

Instituut voor Theoretische Fysica,\\
Universiteit van Amsterdam,\\
Valckernierstraat 65, 1018 XE Amsterdam, The Netherlands\\

\vspace{24pt}

{\sl J. Jurkiewicz}\footnote{Permanent address: Institute of Physics,
Jagellonian University, ul. Reymonta 4, PL-30 059, Krak\'{o}w 16, Poland.
Partly supported by the KBN grant 2PO3B 196 02.}

\vspace{6pt}

Laboratoire de Physique Th\'{e}orique et Hautes Energies\footnote{
Laboratoire associ\'{e} au Centre National de la Recherche Scientifique
-- URA D0063}\\
Universit\'{e} de Paris XI, b\^{a}timent 211, 91405 Orsay Cedex,
France\\
\end{center}
\vspace{24pt}

\vfill

\begin{center}
{\bf Abstract}
\end{center}

\vspace{12pt}

\noindent
A recently proposed renormalization group technique, based on the 
hierarchical structures present in theories with fluctuating 
geometry, is implemented in the model of branched polymers.
The renormalization group equations can be solved analytically,
and the flow in coupling constant space can be determined.

\vfill

\newpage

\section{Introduction}

The renormalization group is one of the most important concepts
in field theory and in the theory of critical phenomena.
It uses two important concepts: ``blocking'' of regions
of space and ``blocked fields''  which are defined in terms
of the fields in the original regions.

In theories like quantum gravity where the fields themselves determine
space the blocking of regions of space becomes a non-trivial notion.
A variety of methods have been proposed \cite{adfj,jkk,bkk,renken,rck,tc}. 
Here we will study
a model for which the methods proposed in \cite{jkk} and further
developed in \cite{bkk} can be implemented and the corresponding
renormalization group equations solved. The model is the so-called
branched polymer model \cite{adfo,adj,bialas}. While this model
indeed is a toy model, it is an appropriately chosen toy model: Whenever
models of fluctuating geometry are studied, branched polymers appear
naturally unless coupling constants are chosen with care.
In two-dimensional quantum gravity, viewed as a statistical model
via dynamical triangulations, it is generally believed that 
the theory degenerates to a theory of branched polymers if 
the central charge $c$ of the matter fields coupled to two-dimensional 
gravity is sufficiently large. Maybe it even happens for all values 
of $c >1$. In three- and four-dimensional quantum gravity, again studied
via dynamical triangulations,  it is known that the weak coupling
phase, i.e. the phase with a small bare gravitational coupling constant,
is a phase of branched polymers \cite{aj,aj1}. Since two- and
four-dimensional quantum gravity, implemented via dynamical
triangulations,  are precisely the models where renormalization
group techniques have been applied, it seems appropriate to study
the method used in \cite{bkk} for branched polymers.

\section{The BP-model}

Let us define the model of branched polymers (the BP-model).

We consider the ensemble of {\em planar rooted planted trees}. A
{\em tree} is a graph without  closed loops. A {\em rooted} tree is
a tree with one marked vertex and {\em planted} tree is a tree
where the {\it degree} of the root (i.e. number of branches) is one.
We call the marked vertex and the link emerging from it
{\it the root} of the planted tree.
Two planted trees are  considered as distinct if they
cannot be mapped on each other by a
continuous deformation of the plane such that the root of one
tree is mapped to the root of the other tree.

The grand canonical partition function is defined as
\beq\label{1.1}
Z(t) =
\sum_{T \in \T} \rho(T),
\eeq
where $\T$ denotes the ensemble of rooted, planted trees and
\beq\label{1.2}
\rho(T) = t_0^{n_0}t_1^{n_1}\dots t_k^{n_k}\dots
\eeq
The $t_i$'s can be viewed as the weights given to
vertices of order $i+1$ relative to the root vertex which
by definition has weight 1, i.e. we have
\beq\label{Z}
Z(t) = t_0 + t_0t_1 + t_0t_1^2 + \cdots + t_0^2 t_2 + \cdots
\eeq
We will consider the $t_i$'s as
our coupling constants and further assume $t_i \geq 0$, although there
exist so called multicritical BP-models where some weights can be negative
\cite{adj}.

The partition function $Z$ satisfies the known equation (see \cite{adj}):
\beq \label{1.3}
Z = \F (Z),
\eeq     
where
\beq \label{1.4}
\F (Z) = \sum_{i=0}^{\infty}t_i Z^i.
\eeq
Sometimes it will be convenient to emphasize that $\F(Z)$ can
be viewed as a function of both $t$ and $Z$.
We will in these situations write $\F(t;Z)$.
We note the following properties. The average number of vertices
of type $i$ above the root is given by
\beq\label{avni}
\la N_i \ra = \frac{t_i}{Z} \frac{d Z }{d t_i},
\eeq
and if we denote by $N$ the total number of vertices above the root we get
from \rf{1.4}
\beq\label{av}
\la N \ra = \frac{1}{1-\F'(Z)},~~~~\la N_i \ra = t_i Z^{i-1} \la N \ra.
\eeq
It is seen that the critical behavior is governed by the function
\beq\label{R}
R(x) = 1-\F'(x) = 1-t_1 -2 t_2 x -3 t_3 x^2 - \cdots
\eeq
The equations
\beq\label{CR}
R(Z(t)) = 0,~~~~Z(t)=\F(Z(t)),
\eeq
where $Z(t) \geq 0$ and $t_i \geq 0$
will determine a critical surface in the space of coupling constants $t_i$
in the sense that the ``volume'' $\la N \ra$ diverges when
$R(Z(t)) \to 0$ for certain values of the coupling constant.

It is easily shown that $\F'(Z)$ has the interpretation as the generating
function for branched polymers with a marked vertex of order 1
a link distance one from the root,
and by iteration that $(\F'(Z))^n$ is the generating
function for branched polymers with one marked vertex
of order 1 separated a link distance $n$ from the root. This
function can be viewed as the two-point function $G(n)$ in the
fractal geometry generated by the branched polymers \cite{adj} and we have
\beq\label{tpf}
G(n) = e^{n \log \cF'(Z)} \to e^{-R(Z) n}
\eeq
for $R(Z) \to 0$. It is seen that $R(Z)$ acts like a mass, and the
scaling limit is again controlled by  $R(Z) \to 0$.

A generic point on the critical surface will correspond to branched polymers.
In fact, let us consider a critical point $t^c_i$ away from
the $t_1=1$ boundary of the
critical surface and introduce the notation $Z_c=Z(t^c)$.
It follows that $Z_c >0$ and $R'(Z_c) < 0$ for a solution to $R(Z_c) =0$.
Let us (for simplicity)
approach the critical point from a point $t$ away from
the critical surface by a deformation $t_1 = t_1^c +\Delta t_1$,
and $t_n = t^c_n$ for $n \neq 1$.
To lowest order we have
\beq\label{gamma}
\oh R'(Z_c) (\Delta Z)^2 +\Delta t_1 Z_c^i =0,
\eeq
which shows that $\Delta Z$ goes like $\sqrt{ \Delta t_1}$, the critical
behavior which characterizes branched polymers.

The structure of the critical surface becomes more transparent
if we note that it follows from the definition of $Z(t)$
and \rf{1.3}-\rf{1.4} that we can write
\beq\label{bpj1}
Z(t) = \frac{t_0}{1-t_1} H\Bigl(\tilde{t}_2,\tilde{t}_3,\ldots,\Bigr),
\eeq
\beq\label{bpj2}
R(Z) = (1-t_1)K(\tilde{t}_2,\tilde{t}_3,\ldots)
\eeq
where
\beq\label{bpj3}
\tilde{t}_n = \frac{t_n t_0^{n-1}}{(1-t_1)^n},~~~~~~n \geq 2,
\eeq
\beq\label{bpj3a}
K(\tilde{t})=\Bigl(1-2\tilde{t}_2 H(\tilde{t})-3\tilde{t}_3 H^2(\tilde{t}) -
\cdots \Bigr)
\eeq
and where $H(\tilde{t}_2,\tilde{t}_3,\ldots)$ has a convergent
power expansion in $\tilde{t}_i$'s around $\tilde{t}_i=0$, with
$H(0,0,\ldots)=1$.

Eq.\ \rf{bpj2} shows that critical behavior (i.e. $R(Z) \to 0$) can
be obtained if either $K(\tilde{t}) \to 0$ or  $t_1\to 1$.
In the first case  we have the branched polymer situation
shown in \rf{gamma}. If $t_1 \to 1$ while the $\tilde{t}_n$'s are constant
we have a new kind of critical behavior, which we denote {\it weakly
branched polymers}. In the simplest case where all $t_n =0$ for $n >1$
\beq\label{Zt1}
Z(t) = \frac{t_0}{1-t_1},
\eeq
i.e. the critical behavior of a linear chain (a polymer) with a chemical
potential $t_1$ per chain link (the vertices of order two).
When $t_1 \to 1$ for constant $\tilde{t}_n$'s we have
\beq\label{bpj4}
\la N \ra = \frac{1}{K(\tilde{t})\, (1-t_1)},
\eeq
while $\la N_n\ra$ stay finite in the limit $t_1 \to 1$ for $n \neq 1$:
\beq\label{bpj5}
\la N_0 \ra = \frac{1}{H(\tilde{t}) K(\tilde{t})},~~~~~~
\la N_n \ra = \frac{\tilde{t}_n H^{n-1} (\tilde{t})}{ K(\tilde{t})}.
\eeq
This limit has an interpretation in terms of polymers which are allowed
to branch and break. Let $p_0$ be the probability per unit polymer length of
breaking, $p_2,p_3,\ldots$ the probability per unit length of
branching in  $3,4,\ldots$ linear pieces.
If the polymers are made of pieces of length $a$ we have
\beq\label{bpj6}
t_0 = a p_0,~~~t_2=ap_2, \ldots, ~~~~1-t_1 = \sum_{i\neq 1} t_i.
\eeq
It is seen that the limit $a \to 0$ corresponds to 
weakly branched polymers since the $\tilde{t}_n$'s are
only functions of the $p_i$'s.

Eq.\ \rf{bpj6} describes the approach to the point $t=(0,1,0,0,\ldots)$.
but the structure \rf{bpj2} is more general.
Assume that the $t_k$'s, $k >1$ are non-negative constants and
let $n >1$ be the smallest integer for which $t_n >0$, assuming
it exists. In this case
only $\tilde{t}_n$ survives in the limit $t_1 \to 1$ and
\beq\label{alpha1}
t_0  =  c(1-t_1)^\frac{n}{n-1}.
\eeq
\begin{figure}
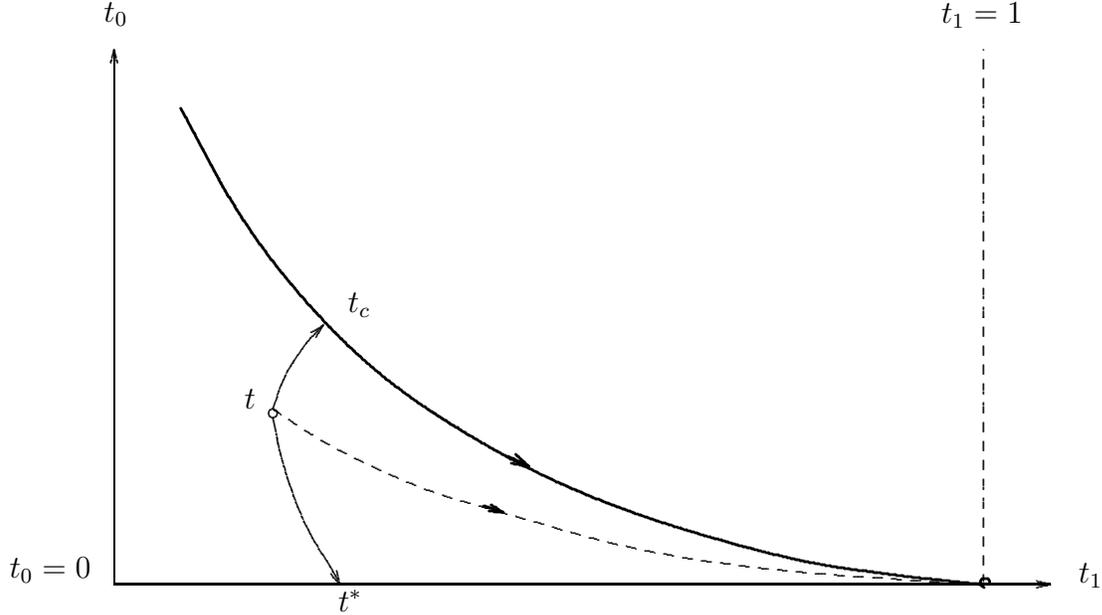

\font\thinlinefont=cmr5
\begingroup\makeatletter\ifx\SetFigFont\undefined
\def\x#1#2#3#4#5#6#7\relax{\def\x{#1#2#3#4#5#6}}%
\expandafter\x\fmtname xxxxxx\relax \def\y{splain}%
\ifx\x\y   
\gdef\SetFigFont#1#2#3{%
  \ifnum #1<17\tiny\else \ifnum #1<20\small\else
  \ifnum #1<24\normalsize\else \ifnum #1<29\large\else
  \ifnum #1<34\Large\else \ifnum #1<41\LARGE\else
     \huge\fi\fi\fi\fi\fi\fi
  \csname #3\endcsname}%
\else
\gdef\SetFigFont#1#2#3{\begingroup
  \count@#1\relax \ifnum 25<\count@\count@25\fi
  \def\x{\endgroup\@setsize\SetFigFont{#2pt}}%
  \expandafter\x
    \csname \romannumeral\the\count@ pt\expandafter\endcsname
    \csname @\romannumeral\the\count@ pt\endcsname
  \csname #3\endcsname}%
\fi
\fi\endgroup
\mbox{\beginpicture
\setcoordinatesystem units <0.700000cm,0.700000cm>
\unitlength=0.70000cm
\linethickness=1pt
\setplotsymbol ({\makebox(0,0)[l]{\tencirc\symbol{'160}}})
\setshadesymbol ({\thinlinefont .})
\setlinear
%
%
\linethickness=1pt
\setplotsymbol ({\makebox(0,0)[l]{\tencirc\symbol{'160}}})
\setdashes < 0.2858cm>
\ellipticalarc axes ratio  0.089:0.089  360 degrees 
	from 19.171  8.924 center at 19.082  8.924
%
%
\linethickness= 0.500pt
\setplotsymbol ({\thinlinefont .})
\setsolid
\putrule from  2.540  8.890 to  2.540 19.050
%
%
\plot  2.603 18.796  2.540 19.050  2.477 18.796 /
%
%
%
\linethickness= 0.500pt
\setplotsymbol ({\thinlinefont .})
\putrule from  2.540  8.890 to 20.320  8.890
%
%
\plot 20.066  8.827 20.320  8.890 20.066  8.954 /
\linethickness= 0.500pt
\setplotsymbol ({\thinlinefont .})
\setdashes < 0.1270cm>
%
%
%
\plot	19.050  8.890 19.050 13.970
 	 /
\plot 19.050 13.970 19.050 19.050 /
\linethickness= 0.500pt
\setplotsymbol ({\thinlinefont .})
%
%
%
\plot	 5.556 12.224  6.032 11.906
 	 6.152 11.828
	 6.271 11.752
	 6.390 11.679
	 6.509 11.609
	 6.628 11.540
	 6.747 11.475
	 6.866 11.411
	 6.985 11.351
	 7.105 11.291
	 7.228 11.232
	 7.353 11.172
	 7.417 11.142
	 7.481 11.113
	 7.546 11.083
	 7.611 11.053
	 7.677 11.023
	 7.744 10.993
	 7.811 10.964
	 7.879 10.934
	 7.948 10.904
	 8.017 10.874
	 8.086 10.845
	 8.156 10.816
	 8.225 10.788
	 8.295 10.760
	 8.364 10.733
	 8.434 10.707
	 8.503 10.681
	 8.572 10.656
	 8.642 10.632
	 8.711 10.608
	 8.781 10.584
	 8.850 10.562
	 8.920 10.540
	 8.989 10.518
	 9.059 10.498
	 9.128 10.478
	 9.198 10.458
	 9.267 10.438
	 9.336 10.418
	 9.406 10.398
	 9.475 10.378
	 9.545 10.358
	 9.614 10.339
	 9.684 10.319
	 9.753 10.299
	 9.823 10.279
	 9.892 10.259
	 9.962 10.239
	10.031 10.220
	10.100 10.200
	10.170 10.180
	10.239 10.160
	10.309 10.140
	10.378 10.120
	10.448 10.100
	10.517 10.081
	10.587 10.061
	10.656 10.041
	10.726 10.021
	10.795 10.001
	10.864  9.981
	10.934  9.962
	11.003  9.942
	11.073  9.922
	11.142  9.902
	11.212  9.882
	11.281  9.862
	11.351  9.842
	11.420  9.823
	11.491  9.803
	11.561  9.784
	11.632  9.765
	11.704  9.746
	11.776  9.728
	11.849  9.710
	11.922  9.692
	11.996  9.674
	12.070  9.656
	12.145  9.639
	12.220  9.622
	12.295  9.605
	12.372  9.589
	12.448  9.573
	12.525  9.557
	12.602  9.541
	12.677  9.526
	12.750  9.511
	12.822  9.497
	12.892  9.484
	12.961  9.470
	13.028  9.458
	13.093  9.446
	13.219  9.423
	13.338  9.402
	13.451  9.383
	13.557  9.366
	13.664  9.350
	13.779  9.334
	13.900  9.316
	13.964  9.308
	14.030  9.299
	14.097  9.290
	14.166  9.281
	14.237  9.271
	14.310  9.262
	14.385  9.253
	14.462  9.243
	14.540  9.233
	14.621  9.223
	14.703  9.214
	14.785  9.204
	14.869  9.194
	14.954  9.185
	15.039  9.176
	15.126  9.167
	15.213  9.158
	15.301  9.149
	15.391  9.140
	15.481  9.132
	15.572  9.123
	15.664  9.115
	15.757  9.107
	15.851  9.099
	15.945  9.091
	16.041  9.084
	16.138  9.076
	16.238  9.068
	16.339  9.061
	16.443  9.053
	16.548  9.045
	16.656  9.038
	16.766  9.030
	16.878  9.022
	16.992  9.014
	17.109  9.006
	17.227  8.998
	17.348  8.990
	17.470  8.981
	17.595  8.973
	17.722  8.965
	17.786  8.961
	17.851  8.957
	17.977  8.949
	18.093  8.941
	18.202  8.934
	18.301  8.928
	18.392  8.922
	18.474  8.916
	18.547  8.911
	18.611  8.907
	18.714  8.899
	18.782  8.894
	18.812  8.890
	 /
\plot 18.812  8.890 18.733  8.890 /
\linethickness=1pt
\setplotsymbol ({\makebox(0,0)[l]{\tencirc\symbol{'160}}})
\setsolid
%
%
%
\plot	 3.810 17.939  4.445 16.748
 	 4.485 16.674
	 4.526 16.600
	 4.567 16.526
	 4.609 16.452
	 4.651 16.378
	 4.694 16.304
	 4.738 16.231
	 4.782 16.158
	 4.827 16.085
	 4.873 16.012
	 4.919 15.939
	 4.966 15.866
	 5.013 15.794
	 5.061 15.722
	 5.110 15.649
	 5.159 15.577
	 5.209 15.505
	 5.260 15.434
	 5.311 15.362
	 5.363 15.291
	 5.415 15.220
	 5.468 15.149
	 5.522 15.078
	 5.576 15.007
	 5.631 14.936
	 5.686 14.866
	 5.743 14.795
	 5.799 14.725
	 5.857 14.655
	 5.915 14.585
	 5.973 14.516
	 6.032 14.446
	 6.092 14.377
	 6.152 14.309
	 6.211 14.241
	 6.271 14.175
	 6.330 14.109
	 6.390 14.043
	 6.449 13.979
	 6.509 13.915
	 6.628 13.790
	 6.747 13.669
	 6.866 13.550
	 6.985 13.434
	 7.104 13.322
	 7.223 13.212
	 7.342 13.106
	 7.461 13.003
	 7.580 12.902
	 7.699 12.805
	 7.818 12.711
	 7.938 12.621
	 8.058 12.531
	 8.182 12.442
	 8.309 12.353
	 8.373 12.308
	 8.439 12.263
	 8.505 12.219
	 8.572 12.174
	 8.639 12.129
	 8.708 12.085
	 8.777 12.040
	 8.847 11.996
	 8.918 11.951
	 8.989 11.906
	 9.062 11.862
	 9.135 11.817
	 9.209 11.772
	 9.283 11.728
	 9.359 11.683
	 9.435 11.638
	 9.512 11.594
	 9.589 11.549
	 9.668 11.504
	 9.747 11.460
	 9.827 11.415
	 9.908 11.370
	 9.990 11.326
	10.072 11.281
	10.155 11.237
	10.239 11.192
	10.324 11.147
	10.410 11.103
	10.497 11.059
	10.584 11.016
	10.673 10.973
	10.762 10.930
	10.853 10.887
	10.944 10.845
	11.036 10.803
	11.129 10.761
	11.223 10.720
	11.318 10.678
	11.414 10.638
	11.511 10.597
	11.609 10.557
	11.708 10.517
	11.807 10.478
	11.908 10.438
	12.010 10.400
	12.112 10.361
	12.216 10.323
	12.320 10.285
	12.425 10.247
	12.531 10.210
	12.638 10.173
	12.747 10.136
	12.855 10.099
	12.965 10.063
	13.076 10.027
	13.188  9.992
	13.301  9.957
	13.414  9.922
	13.528  9.888
	13.640  9.854
	13.750  9.821
	13.860  9.789
	13.967  9.758
	14.074  9.727
	14.179  9.698
	14.283  9.669
	14.385  9.641
	14.486  9.613
	14.585  9.587
	14.683  9.561
	14.780  9.536
	14.875  9.512
	14.969  9.488
	15.061  9.465
	15.152  9.444
	15.242  9.422
	15.330  9.402
	15.417  9.382
	15.503  9.364
	15.587  9.345
	15.670  9.328
	15.751  9.312
	15.831  9.296
	15.909  9.281
	15.987  9.267
	16.062  9.253
	16.137  9.241
	16.210  9.229
	16.281  9.218
	16.351  9.207
	16.420  9.198
	16.489  9.188
	16.557  9.178
	16.624  9.169
	16.691  9.160
	16.757  9.151
	16.822  9.142
	16.887  9.133
	16.951  9.124
	17.015  9.116
	17.140  9.100
	17.263  9.084
	17.383  9.069
	17.501  9.054
	17.616  9.040
	17.729  9.027
	17.840  9.014
	17.947  9.002
	18.053  8.990
	18.156  8.980
	18.256  8.969
	 /
\plot 18.256  8.969 19.050  8.890 /
\linethickness= 0.500pt
\setplotsymbol ({\thinlinefont .})
%
%
%
\plot	 5.556 12.065  5.636 11.668
 	 5.657 11.568
	 5.680 11.465
	 5.706 11.359
	 5.735 11.251
	 5.766 11.141
	 5.799 11.028
	 5.835 10.913
	 5.874 10.795
	 5.913 10.678
	 5.953 10.567
	 5.993 10.460
	 6.032 10.358
	 6.072 10.262
	 6.112 10.170
	 6.152 10.083
	 6.191 10.001
	 6.231  9.923
	 6.271  9.847
	 6.310  9.774
	 6.350  9.704
	 6.390  9.635
	 6.429  9.570
	 6.509  9.446
	 6.583  9.332
	 6.648  9.227
	 6.702  9.133
	 6.747  9.049
	 /
\plot  6.747  9.049  6.826  8.890 /
%
%
\plot  6.656  9.089  6.826  8.890  6.769  9.146 /
\linethickness=1pt
\setplotsymbol ({\makebox(0,0)[l]{\tencirc\symbol{'160}}})
\setdashes < 0.2858cm>
%
%
%
\plot	 9.557 10.389  9.739 10.325
 	 /
\plot  9.739 10.325  9.921 10.262 /
\setsolid
%
%
\plot  9.660 10.285  9.921 10.262  9.702 10.405 /
\setdashes < 0.2858cm>
%
%
\linethickness= 0.500pt
\setplotsymbol ({\thinlinefont .})
\setsolid
\ellipticalarc axes ratio  0.080:0.080  360 degrees 
	from  5.637 12.145 center at  5.556 12.145
\linethickness=1pt
\setplotsymbol ({\makebox(0,0)[l]{\tencirc\symbol{'160}}})
\setdashes < 0.2858cm>
%
%
%
\plot	10.033 11.326 10.224 11.231
 	 /
\plot 10.224 11.231 10.414 11.136 /
\setsolid
%
%
\plot 10.087 11.193 10.414 11.136 10.173 11.363 /
\setdashes < 0.2858cm>
%
%
\put{\SetFigFont{12}{14.4}{it}$t_1=1$} [B] at 19.018 19.560
\linethickness= 0.500pt
\setplotsymbol ({\thinlinefont .})
\setsolid
%
%
%
\plot	 5.556 12.224  5.636 12.462
 	 5.675 12.576
	 5.715 12.680
	 5.755 12.774
	 5.794 12.859
	 5.839 12.943
	 5.894 13.037
	 5.958 13.142
	 6.032 13.256
	 6.112 13.370
	 6.191 13.474
	 6.271 13.568
	 6.350 13.652
	 /
\plot  6.350 13.652  6.509 13.811 /
%
%
\plot  6.374 13.587  6.509 13.811  6.284 13.677 /
%
%
%
\put{\SetFigFont{12}{14.4}{it}$t$} [lB] at  5.017 12.226
%
%
\put{\SetFigFont{12}{14.4}{it}$t^*$} [lB] at  6.794  8.384
%
%
\put{\SetFigFont{12}{14.4}{it}$t_c$} [lB] at  6.985 14.067
%
%
\put{\SetFigFont{12}{14.4}{it}$t_1$} [lB] at 20.860  8.924
%
%
\put{\SetFigFont{12}{14.4}{it}$t_0=0$} [rB] at  2.095  9.019
%
%
\put{\SetFigFont{12}{14.4}{it}$t_0$} [B] at  2.572 19.592
\linethickness=0pt
\putrectangle corners at  1.118 19.897 and 20.860  8.327
\endpicture}

\caption{The phase $t_0,t_1$ part of the phase diagram.
The thick line symbolizes the critical surface for fixed $t_n$, $n >1$.
$t$ is the initial choice of coupling constants.
The flow to $t^*$ is obtained by the
renormalization group transformation and corresponds to
a flow to a finite linear chain.
The approach to $t_c$ at the critical surface  leads
to the generic BP while the approach to the critical surface along
the dashed line leads to the weakly branched polymers. Finally, the
arrow at the critical surface indicates the flow {\it on} the
critical surface towards $t_1=1$ under the action of the
renormalization group transformations. }
\end{figure}
This implies that $Z(t) \to 0$ and
\beq\label{t1}
\la N_0 \ra \to {\rm const},~~~\la N_n \ra \to {\rm const}~~~~~~~
\la N_1 \ra \to \frac{1}{1-t_1},
\eeq
while all other $\la N_i \ra$'s are either identically zero or approach
zero for $t_1 \to 1$. 

The different ways of approaching the critical surface
are illustrated in fig.\ 1. It is of course possible to 
fine-tune the approach to the critical surface such that 
both $t_1 \to 1$ and $K(\tilde{t}) \to 0$ and obtain a
hybrid between weakly branched polymers and ordinary branched 
polymers. Note the different role of the coupling constants $t_i$, $i \geq 2$
at the boundary $t_1=1$ and for $t_1 < 1$ at the critical surface.

\section{RG equations}

In  \cite{jkk,bkk} the following procedure was suggested as a
replacement for the conventional blocking of cells in real space
renormalization group transformations in the case of fluctuating
geometry: cut away the last generation of baby universe outgrowths,
i.e. baby universes which are not themselves contained in any
baby universes. This was viewed as a way to cut away fine structure
details and in a general model with many coupling constants the
new fractal structure should be obtained by a change in coupling constants.
In this way successive cutting should induce a renormalization group
flow in the coupling constant space.

Let us apply the construction to the BP-model.
The vertices can be  classified as  {\it outer vertices},
i.e. the vertices of order one, and  {\it inner vertices}.
The marked vertex at the
root has a special status and is not
allowed to be touched.  The renormalization group transformation consists
of cutting away all outer vertices and their associated links.
Every rooted tree can be obtained from another tree by this procedure,
but they will have a new weight
\beq \label{2.1}
\rho^{(1)}(\T ) = \sum_{\T ' : RG(\T ')=\T} \rho^{(0)}(\T ')
\eeq
where the sum is over all trees $\T '$
which after the transformation coincide
with $\T$. The trees $\T '$ are generated by taking a tree $\T $ and
i) adding one or more outer vertices (and associated links)
to all outer vertices in $T$ (which then cease to be outer vertices in $T'$)
and ii) adding zero or more outer vertices (and associated links)
as neighbors to  the inner vertices.
Notice that the trees $T'$ will always be different from
the {\em root term} consisting of a single link with
two vertices,  the marked one  and an outer vertex. This means that
$\rho^{(1)}$ will not contain the term $t_0$.

It is not too difficult to convince oneself that
\beq \label{2.2}
\rho^{(1)}(\T) = (t_0^{(1)}/t_0)^{n_0}(t_1^{(1)}/t_1)^{n_1}\dots
\rho^{(0)}(\T),
\eeq
which corresponds to the following
redefinition of the coupling constants $t_i$
\bea
t_0^{(1)} & = & \F(t;x)|_{x=t_0} - t_0 \nonumber  \\
t_l^{(1)} & = & \frac{1}{l!}\frac{\partial^l \F(t;x)}{\partial x^l}
|_{x = t_0},~~~l>0. \label{2.4}
\eea

These redefinitions can be put in a RG equation:
\beq \label{2.5}
Z(\{ t_i \}) - t_0 = Z(\{ t_i^{(1)} \}).
\eeq

Observe that the transformation (\ref{2.4}) corresponds to the
following redefinition of $Z$:
\bea
Z^{(1)} & = & Z - t_0, \nonumber\\
Z^{(1)} & = & \F(t;x)|_{x=t_0 + Z^{(1)}} - \F(t;x)|_{x=t_0}
= \F^{(1)}(Z^{(1)}),\label{2.6}
\eea
where $F^{(1)}(Z^{(1)})$ is defined as $\F(Z)$, just using $t^{(1)}$ and
$Z^{(1)}$ variables. It follows by definition that
\beq\label{inv}
\F'(t;x)|_{x=Z} = {\F^{(1)}}'(t;x)|_{x=Z^{(1)}},
\eeq
where differentiation  is with respect to $x$.
According to \rf{av}  $\la N \ra$ depends only on $\F'(x)$ and 
is consequently an invariant
under the renormalization group transformation.

Equations (\ref{2.6}) will be the central point of this discussion. They
can be used to study the flow of the coupling constants $t_i$ under the
action of RG. We can iterate (\ref{2.6}) and obtain the following recursion
relation
\bea
\F^{(k+1)}(x) & = & \F^{(k)}(\F^{(k)}(0)+x)-\F^{(k)}(0),\nonumber\\
\F^{(0)}(x) & = & \F(x). \label{2.7}
\eea
The corresponding changes in $Z$ are
\bea
Z^{(k+1)} & = & Z^{(k)} - \F^{(k)}(0),\nonumber \\
Z^{(0)} & = & Z.\label{2.8}
\eea
where $Z^{(k)}$ satisfies
\beq \label{2.9}
Z^{(k)}= \F^{(k)}(Z^{(k)}).
\eeq
Equations (\ref{2.8}) and (\ref{2.9}) can be solved
\bea
\F^{(k)}(x) &=&\F(\alpha_k+x)-\alpha_k,\nonumber\\
\alpha_{k+1} &=& \F(\alpha_k),\label{2.10}  \\
\alpha_0 & = & 0.\nonumber
\eea
Similarly
\beq \label{2.11}
Z^{(k)}=Z - \alpha_k.
\eeq

The fixed point of these equations is
\bea
\F^{(k)}(x) &\to& \F^*(x)=\F(\alpha^*+x)-\alpha^*,\nonumber\\
Z^{(k)} & \to & Z^* = Z - \alpha^*,\label{2.12}
\eea
where $\alpha^*$ satisfies
\beq \label{2.13}
\alpha^* = \F(\alpha^*).
\eeq

If we start out with an arbitrary set of (non-negative) coupling constants
$t_i$ the partition function will either be divergent
($Z = \F(Z)$ has no solution) or it will have one or two solutions.
In case it has one solution $R(Z) =0$ and we are at the critical
surface, but generically it will have two solutions, of which we should
choose the one with the smallest value of $Z$. Applying the
transformation group the coupling constants will flow to $t^*$,
characterized by $\a^*$. However,  we know from \rf{2.13} and \rf{1.3} that
\beq\label{surp}
\a^* = Z(t).
\eeq
In other words
\bea
Z^* & =& 0,\nonumber\\
\F^*(x) &=& \sum_{l=1}^{\infty}t_l^* x^l, \label{2.14}\\
t_0^* &=& 0,\nonumber
\eea
where the coefficients $t_l^*$ are the
Taylor expansion coefficients of $\F(x)$
around a point $x=\alpha^*=Z(t)$
\beq \label{2.15}
t_l^* = \frac{1}{l!}\frac{\partial^l \F(x)}{\partial x^l}|_{x=\alpha*}.
\eeq
The $k \to \infty$ limit of (\ref{2.9}) and (\ref{2.10}) can be described
by putting
\beq \label{2.16}
\alpha_k = \alpha^* - \e_k.
\eeq
Close to the fixed point we get from $\a_{k+1} = \F(\a_k)$
\beq\label{approach}
\ep_{k+1} \approx \F'(\a^*) \ep_k = \Bigl(1- \frac{1}{\la N \ra} \Bigr) \e_k.
\eeq
In addition we have
\bea
Z^{(k)} &=& \e_k \label{2.17}\\
t_0^{(k)} &=& \F(\alpha_k)-\alpha_k \approx \frac{\e_k}{\la N \ra},\nonumber
\eea
and in effect
\beq \label{2.18}
\la N^{(k)}_0 \ra =\frac{t_0^{(k)}}{Z^{(k)}}\la N \ra \to 1,
\eeq
for $k \to \infty$. It is easy to check that
\beq\label{lim}
\la N_i^{(k)} \ra \to 0,
\eeq
for $i >1$ since $Z^{(k)} \to 0$, while $t^*_l$ and $\la N \ra$ are fixed
under repeated application of the renormalization transformation.
It is clear that we simply end up with a linear chain length $\la N \ra$.
The flow is shown in fig.\ 1.

\section{Discussion}

The flow of coupling constants under the action of the
renormalization group transformation used here is such that we move
{\it away} from the critical line corresponding to branched polymers
and directly to a linear chain of the same volume. The usual
situation when applying the renormalization group transformations
starting at a point in coupling constant space close to the
critical surface is that one moves towards the critical surface
in the first couple of steps since the irrelevant operators will
dominate the blocking in the first few iterations. Eventually, after
repeated applications of the renormalization group transformations
one moves away from the critical surface in the direction
dictated by the most relevant operator. In the BP-model all coupling
constants $t_n$, $n > 1$ correspond to relevant couplings and
repeated application immediately move us away from the critical surface.
The only trace of being close to critical surface is that 
the approach to $t^*$ will be slower as $\la N \ra$ increases.
If we place ourself {\it at}
the critical surface the iteration of the renormalization group
will not remove us away from this surface and we will
move towards the critical point $t^*_1=1$. The approach to $t^*$
will not be exponentially fast in the number of iterations $k$,
as was the case
for finite $\la N \ra$ in eqs.\ \rf{approach} and \rf{2.17}. Rather,
it is replaced by a power approach in $k$:
\beq\label{power}
\ep_k \approx \frac{2}{-R'(\a^*)k},~~~~t^{(k)}_0 \approx -\oh R'(\a^*) \ep_k
\eeq
which results in the following behavior
\beq\label{final}
\frac{\la N^{(k)}_0 \ra}{\la N\ra} \approx \frac{1}{k},
~~~~ \frac{\la N^{(k)}_2 \ra}{\la N\ra} \approx \frac{1}{k},
~~~~\frac{\la N^{(k)}_i \ra}{\la N\ra} = O(({1}/{k})^{i-1}).
\eeq
Approaching the critical point $t^*_1=1$ the branched polymer becomes
more and more like a weakly branched polymer of the simplest kind,
i.e. corresponding to $n=2$.

\section{Acknowledgments}
P.B. would like to thank Zdzislaw Burda and Andrzej Krzywicki for many
helpful discussions and comments and the 'Stichting voor Fundamenteel
Onderzoek der Materie' (FOM) for financial support.

\end{document}